\documentclass{article}

\usepackage{subcaption}
\usepackage{graphicx}
\usepackage{xurl}
\usepackage{amssymb}
\usepackage{amsmath}
\usepackage{algorithm,algorithmic}
\usepackage{multicol}
\usepackage{multirow}
\usepackage{color}
\usepackage{authblk}

\begin{document}
\title{Accelerating Nonlinear Time-History Analysis with Complex Constitutive Laws via Heterogeneous Memory Management: From 3D Seismic Simulation to Neural Network Training}

\author[1]{Tsuyoshi Ichimura}
\author[1,2]{Kohei Fujita}
\author[1]{Hideaki Ito}
\author[3]{Muneo Hori}
\author[1]{Maddegedara Lalith}
\affil[1]{Earthquake Research Institute and Department of Civil Engineering, The University of Tokyo, Japan}
\affil[2]{RIKEN Center for Computational Science, Japan}
\affil[3]{Research Institute for Value-Added-Information Generation, Japan Agency for Marine-Earth Science and Technology, Japan}

\date{}
\maketitle

\begin{abstract}
Nonlinear time-history evolution problems employing high-fidelity physical models are essential in numerous scientific domains. However, these problems face a critical dual bottleneck: the immense computational cost of time-stepping and the massive memory requirements for maintaining a vast array of state variables. To address these challenges, we propose a novel framework based on heterogeneous memory management for massive ensemble simulations of general nonlinear time-history problems with complex constitutive laws. Taking advantage of recent advancements in CPU-GPU interconnect bandwidth, our approach actively leverages the large capacity of host CPU memory while simultaneously maximizing the throughput of the GPU. This strategy effectively overcomes the GPU memory wall, enabling memory-intensive simulations. We evaluate the performance of the proposed method through comparisons with conventional implementations, demonstrating significant improvements in time-to-solution and energy-to-solution. Furthermore, we demonstrate the practical utility of this framework by developing a Neural Network-based surrogate model using the generated massive datasets. The results highlight the effectiveness of our approach in enabling high-fidelity 3D evaluations and its potential for broader applications in data-driven scientific discovery.
\end{abstract}

\section{Introduction}
Nonlinear time-history evolution problems play a pivotal role in the understanding and prediction of phenomena across various scientific domains. While high-fidelity physical models and finer discretization are essential for the faithful simulation of real-world phenomena, the adoption of such models not only increases computational costs but also demands an immense amount of computer memory to maintain a vast array of state variables at each time step, making the simulations highly memory-intensive. In particular, while GPU computational performance has improved dramatically, the capacity limitations of GPU memory present a severe memory wall, acting as a primary bottleneck that hinders the realization of the aforementioned large-scale, high-fidelity simulations.

To address these memory capacity constraints, conventional approaches have relied on simplifying physical models or employing multi-node parallelization using GPUs exclusively without utilizing CPU memory; however, these methods often lead to reduced accuracy of simulations or increased resource costs. On the other hand, while CPU-only approaches -- leveraging the larger memory capacity of host systems -- can alleviate memory shortages, they often suffer from slower processing speeds compared to GPUs, resulting in a degradation of the time-to-solution. In particular, for massive ensemble simulations -- such as those required for uncertainty quantification (UQ) and dataset generation for machine learning, which have seen increasing demand -- this slow computational speed and high resource cost become major drawbacks.

Meanwhile, motivated by recent increases in CPU-GPU interconnect bandwidth (e.g., \cite{PCIe}, \cite{GH200}), research and development in heterogeneous computing are actively pursuing simultaneous improvements in both time-to-solution and energy-to-solution through tighter coupling of CPUs and GPUs (e.g., \cite{waccpd}). In light of these trends and to address the aforementioned challenges, this study proposes a novel framework based on heterogeneous memory management for massive ensemble simulations of general nonlinear time-history problems with complex constitutive laws, which effectively leverages host CPU memory while maximizing GPU throughput in a heterogeneous CPU-GPU environment. In Section~\ref{sct2}, we describe the proposed method using nonlinear seismic ground response analysis as an example and evaluate its performance in terms of time-to-solution and energy-to-solution through comparisons with conventional implementations. In Section~\ref{sct3}, we demonstrate the practical utility of this framework by developing a Neural Network-based surrogate model using the massive datasets generated through these ensemble simulations.

\section{Method}
\label{sct2}

While the proposed method is applicable to general time-history evolution problems, we describe it in detail using 3D nonlinear seismic ground response analysis with complex constitutive laws as a concrete example to provide specificity. To faithfully reproduce real-world ground responses, a detailed 3D finite element model using finer unstructured grids and high-fidelity nonlinear physical constitutive laws is indispensable, resulting in a simulation that requires both immense memory capacity and high-speed computation.

\subsection{Target problem}

First, we describe the 3D nonlinear seismic ground response analysis. Following \cite{asme2014}, taking advantage of the finite element method's superiority in modeling complex geometries and satisfying stress-free boundary conditions at the ground surface, we discretize the nonlinear wave equation -- in which material properties change from moment to moment due to nonlinear soil constitutive laws -- and solve the following equation discretized by the Newmark-$\beta$ method at each time step to obtain the response of the nonlinear time-history problem:
\begin{equation}
\left( \frac{4}{dt^2}\mathbf{M}+\frac{2}{dt}\mathbf{C}^n+\mathbf{K}^n\right)\delta \mathbf{u}^n = \mathbf{f}^n-\mathbf{q}^{n-1}+\mathbf{C}^n\mathbf{v}^{n-1}+\mathbf{M}\left(\mathbf{a}^{n-1}+\frac{4}{dt}\mathbf{v}^{n-1}\right)
\label{eq:GE}
\end{equation}
with
$\mathbf{q}^n=\mathbf{q}^{n-1}+\mathbf{K}^n\delta\mathbf{u}^n$,
$\mathbf{u}^n=\mathbf{u}^{n-1}+\delta\mathbf{u}^n$,
$\mathbf{v}^n=-\mathbf{v}^{n-1}+\frac{2}{dt}\delta\mathbf{u}^n$, and
$\mathbf{a}^n=-\mathbf{a}^{n-1}-\frac{4}{dt}\mathbf{v}^{n-1} +\frac{4}{dt^2}\delta\mathbf{u}^n$.
Here, $\mathbf{M}$, $\mathbf{C}^n$, and $\mathbf{K}^n$ denote the mass matrix and the damping and stiffness matrices at the $n$-th time step, respectively, while $\delta\mathbf{u}^n$, $\mathbf{u}^n$, $\mathbf{v}^n$, and $\mathbf{a}^n$ represent the nodal vectors for displacement increment, displacement, velocity, and acceleration. Furthermore, $dt$ denotes the time step interval. Rayleigh damping is employed for $\mathbf{C}^n$, which is determined using the damping $h^n$ at the $n$-th time step, following the procedure in \cite{asme2014}. Consequently, in Eq.~\eqref{eq:GE}, the matrix equations including $\mathbf{K}^n$ and $\mathbf{C}^n$ must be solved while maintaining a massive number of state variables that evolve at each time step $n$ according to the nonlinear soil constitutive laws. For the finite element discretization, second-order tetrahedral elements are used due to their excellent geometric modeling capabilities and the necessity of strain evaluation for the constitutive laws. Semi-infinite absorbing boundary conditions are applied to the bottom and side boundaries.

To achieve simulations with a higher level of fidelity than \cite{asme2014}, this study adopts the multi-spring model \cite{multispring} as a more sophisticated nonlinear constitutive law. In the finite element method, the element stiffness matrix $\mathbf{K}_e$, which constitutes $\mathbf{K}^n$ in Eq.~\eqref{eq:GE}, is constructed as follows:
\begin{equation}
\mathbf{K}_e=\sum_{j=1}^5 w_j \mathbf{B}^T_{e,j} \mathbf{D}_{e,j} \mathbf{B}_{e,j}
\label{eq:ms}
\end{equation}
Here, $\mathbf{B}^T{e,j}$ is a $6 \times 30$ matrix that converts nodal displacements into strains, $\mathbf{D}_{e,j}$ is a $6 \times 6$ elasto-plastic stiffness matrix at the integration points, and $w_j$ represents the weights of the integration points. In the multi-spring model, $\mathbf{D}$ is determined by combining numerous empirically derived 1D nonlinear springs, which necessitates the storage of a large number of state variables. Specifically, since this study employs the modified Ramberg-Osgood model \cite{RO} and Masing rule \cite{massing} as the constitutive laws for the 1D springs, it is necessary to maintain 40 bytes of data per spring, consisting of four double-precision variables and two flags. Furthermore, with 150 1D springs per material evaluation point and four evaluation points per tetrahedral element, a total of 24 kbytes of data must be stored per element, resulting in a highly memory-intensive simulation.

\subsection{Proposed method}

We propose a method that leverages the characteristics of both the large capacity of CPU memory and the high performance of GPUs for the aforementioned memory-intensive simulations. The target problem in this study consists of two types of computations: a memory-capacity-bound constitutive law calculation part and a compute-intensive linear equation solver part; while the former requires CPU memory because it exceeds GPU memory capacity, the latter is expected to achieve high-speed computation by utilizing the high-bandwidth memory and high-performance cores of the GPU. In conventional computers with limited CPU-GPU interconnect bandwidth (e.g., older PCIe), even if the compute-intensive part is executed on the GPU while keeping the memory-capacity-bound part on the CPU, the data transfer between them becomes a bottleneck, preventing overall speedup and leading to the use of CPU-only execution for both parts; however, in systems with relatively high CPU-GPU bandwidth (e.g., PCIe Gen 5 x16), GPU performance can be utilized by offloading the compute-intensive part to the GPU and performing frequent data transfers between the memory-capacity-bound part. To achieve even higher speeds, we introduce heterogeneous memory management for systems with ultra-high CPU-GPU bandwidth, such as the GH200, enabling all computations to be performed on high-performance GPUs by overlapping GPU computing with CPU-GPU data transfer in the memory-capacity-bound section while keeping massive data in CPU memory. Furthermore, the efficiency of computational performance is enhanced by employing an algorithm that manages the memory hierarchy within the GPU for the compute-intensive part. Details are described below using the target problem as an example.

The primary computations in the target problem are in solving Eq.~\eqref{eq:GE} and the multi-spring calculation in Eq.~\eqref{eq:ms}. Since Eq.~\eqref{eq:GE} involves a large number of degrees of freedom and the target matrix is a positive-definite symmetric sparse matrix, iterative solvers based on the conjugate gradient method are typically employed for its solution. Since sparse matrices are often stored in memory using compressed formats such as compressed row storage (CRS), this solver part corresponds to the compute-intensive component. Conversely, the multi-spring calculation in Eq.~\eqref{eq:ms} requires a massive memory capacity to store variables for numerous 1D springs that evolve over time, corresponding to the memory-capacity-bound part. In this study, as a conventional baseline where all computations are executed on the CPU -- typical for systems with limited CPU-GPU bandwidth -- we utilize a preconditioned conjugate gradient method with $3 \times 3$ block Jacobi preconditioning and CRS-based sparse matrix-vector multiplication (hereafter referred to as "Baseline Method 1: CRSCPU\_MSCPU"; see Algorithm~\ref{alg:CRSCPU_MSCPU}). Additionally, as a baseline for systems with moderate CPU-GPU bandwidth, we define "Baseline Method 2: CRSGPU\_MSCPU" (Algorithm~\ref{alg:CRSGPU_MSCPU}), where the multi-spring calculation is performed on the CPU while the solver part is offloaded to the GPU, with data exchanged between them. In this method, the solver and the CRS updates in Baseline Method 1 are offloaded to the GPU.

\begin{algorithm}[tb]
\caption{\small{Baseline method 1: CRSCPU\_MSCPU. $\mathbf{D}$ indicate the stiffness matrix evaluated by the multispring method, while $\theta$ indicate the nonlinear spring parameters. CRS-PCG indicate $3\times3$ block Jacobi preconditioned conjugate gradient solver with CRS-based matrix-vector products.}}
\label{alg:CRSCPU_MSCPU}
\begin{algorithmic}[1]
\small{
\FOR{$it = 1;~ it \le nt;~ it=it+1$}
\STATE ~~ $\delta \mathbf{u}^{it} \Leftarrow$ CRS-PCG($\mathbf{A}$, $\mathbf{f}^{it}$)@CPU \\
\STATE ~~ $\{ \mathbf{D}^{it}, \theta^{it}\} \Leftarrow$ Multispring($\delta \mathbf{u}^{it}, \theta^{it-1}$)@CPU \\
\STATE ~~ $\mathbf{A} \Leftarrow$ UpdateCRS($\mathbf{D}^{it}$)@CPU \\
\ENDFOR
}
\end{algorithmic}
\end{algorithm}

\begin{algorithm}[tb]
\caption{\small{Baseline method 2: CRSGPU\_MSCPU. $\mathbf{D}$ indicate the stiffness matrix evaluated by the multispring method, while $\theta$ indicate the nonlinear spring parameters. CRS-PCG indicate $3\times3$ block Jacobi preconditioned conjugate gradient solver with CRS-based matrix-vector products.}}
\label{alg:CRSGPU_MSCPU}
\begin{algorithmic}[1]
\small{
\FOR{$it = 1;~ it \le nt;~ it=it+1$}
\STATE ~~ $\delta \mathbf{u}^{it} \Leftarrow$ CRS-PCG($\mathbf{A}$, $\mathbf{f}^{it}$)@GPU \\
\STATE ~~ Transfer $\delta \mathbf{u}^{it}$ from GPU to CPU
\STATE ~~ $\{ \mathbf{D}^{it}, \theta^{it}\} \Leftarrow$ Multispring($\delta \mathbf{u}^{it}, \theta^{it-1}$)@CPU \\
\STATE ~~ Transfer $\mathbf{D}^{it}$ from CPU to GPU
\STATE ~~ $\mathbf{A} \Leftarrow$ UpdateCRS($\mathbf{D}^{it}$)@GPU \\
\ENDFOR
}
\end{algorithmic}
\end{algorithm}

In the proposed method, we achieve high-speed computation by storing variables for the compute-intensive part in GPU memory while performing the memory-capacity-bound calculations on the GPU by accessing data in CPU memory via high-bandwidth transfers, effectively executing the entire simulation on the high-performance GPU. Here, the multi-spring calculation is divided into multiple blocks, and the time required for data transfer is hidden by overlapping the GPU computation of one block with the CPU-GPU transfer of another block, thereby circumventing GPU memory capacity constraints and reducing execution time. As shown in Algorithm~\ref{alg:CRSGPU_MSGPU}, this method pipelines the CPU-GPU transfer of multi-spring data ($\mathbf{\theta}$) (Line 6) and the multi-spring calculation (Line 7). While the entire $npart$ blocks must be stored on the CPU side, the GPU only needs to store two blocks at a time, allowing large-scale problems to be solved within the limited GPU memory. Hereafter, this approach is referred to as "Proposed Method 1: CRSGPU\_MSGPU."

While Proposed Method 1 addresses heterogeneous memory management between CPU and GPU, even more efficient GPU computation can be achieved through effective management of the intra-GPU memory hierarchy (device memory, cache, and registers). We employ the Element-by-Element (EBE) method \cite{EBE}, which reduces memory footprint and transfer size at the cost of increased computational operations for the sparse matrix-vector multiplication in the solver. This converts the CRS-based sparse matrix-vector multiplication, which is typically memory-bandwidth-bound, into an atomic-add-bound operation on the L2 cache, enabling higher computational throughput. Additionally, the reduction in memory usage allows two problem sets to be loaded into GPU memory simultaneously -- whereas CRS-based methods could accommodate only one -- effectively doubling the degrees of freedom per node and reducing random access costs in sparse matrix-vector multiplication. The method optimized through such intra-GPU memory hierarchy management is called as "Proposed Method 2: EBEGPU\_MSGPU\_2SET" (Algorithm~\ref{alg:EBEGPU_MSGPU_2SET}). Furthermore, because surplus GPU memory remains even with two problem sets loaded, we leverage this extra capacity to apply a variable-precision multi-grid preconditioner \cite{SC14} to the solver for even greater efficiency.

\begin{algorithm}[tb]
\caption{\small{Proposed method 1: CRSGPU\_MSGPU. Data $\mathbf{D}_j$ and $\theta_j$ indicate the $j$-th partition of stiffness matrix and the nonlinear spring parameters, respectivey. Lines 6 and 7 are conducted simultaneously, leading to overlap of GPU-computation and CPU-GPU data transfer. While $npart$ partitions of data is required in CPU memory, only 2 partitions reside on GPU memory at once; leading to using GPU computing under the low GPU-memory footprint constraint.}}
\label{alg:CRSGPU_MSGPU}
\begin{algorithmic}[1]
\small{
\FOR{$it = 1;~ it \le nt;~ it=it+1$}
\STATE $\delta \mathbf{u}^{it} \Leftarrow$ CRS-PCG($\mathbf{A}$, $\mathbf{f}^{it}$)@GPU \\
\STATE Transfer $\theta^{it-1}_1$, $\theta^{it-1}_2$ from CPU to GPU
\STATE $\{ \mathbf{D}^{it}_1, \theta^{it}_1\} \Leftarrow$ Multispring($\delta \mathbf{u}^{it}, \theta^{it-1}_1$)@GPU \\
\FOR{$j = 2;~ j \le npart-1;~ j=j+1$}
\STATE Transfer $\theta^{it}_{j-1}$ from GPU to CPU, Transfer $\theta^{it-1}_{j+1}$ from CPU to GPU
\STATE $\{ \mathbf{D}^{it}_j, \theta^{it}_j\} \Leftarrow$ Multispring($\delta \mathbf{u}^{it}, \theta^{it-1}_j$)@GPU \\
\ENDFOR
\STATE $\{ \mathbf{D}^{it}_{npart}, \theta^{it}_{npart}\} \Leftarrow$ Multispring($\delta \mathbf{u}^{it}, \theta^{it-1}_{npart}$)@GPU \\
\STATE Transfer $\theta^{it}_{npart-1}$, $\theta^{it}_{npart}$ from GPU to CPU
\STATE $\mathbf{A} \Leftarrow$ UpdateCRS($\mathbf{D}^{it}$)@GPU \\
\ENDFOR
}
\end{algorithmic}
\end{algorithm}

\begin{algorithm}[tb]
\caption{\small{Proposed method 2: EBEGPU\_MSGPU\_2SET. Note that lines 2 and 3 are conducted together to reduce random memory access costs. Also, lines 4 and 5 are conducted together. Pipelined GPU computation is the same contents as in lines 3--10 in Algorithm~\ref{alg:CRSGPU_MSGPU}. EBE-IPCG indicate an adaptive conjugate gradient solver with mixed precision multigrid-based preconditioner \cite{SC14} with EBE-based matrix-vector products.}}
\label{alg:EBEGPU_MSGPU_2SET}
\begin{algorithmic}[1]
\small{
\FOR{$it = 1;~ it \le nt;~ it=it+1$}
\STATE $\delta \mathbf{u}^{it}_1 \Leftarrow$ EBE-IPCG($\mathbf{D}^{it-1}_1$, $\mathbf{f}^{it}_1$)@GPU \\
\STATE $\delta \mathbf{u}^{it}_2 \Leftarrow$ EBE-IPCG($\mathbf{D}^{it-1}_2$, $\mathbf{f}^{it}_2$)@GPU \\
\STATE Pipelined GPU computation of $\{ \mathbf{D}^{it}_1, \theta^{it}_1\} \Leftarrow$ Multispring($\delta \mathbf{u}^{it}_1, \theta^{it-1}_1$)@GPU \\
\STATE Pipelined GPU computation of $\{ \mathbf{D}^{it}_2, \theta^{it}_2\} \Leftarrow$ Multispring($\delta \mathbf{u}^{it}_2, \theta^{it-1}_2$)@GPU \\
\ENDFOR
}
\end{algorithmic}
\end{algorithm}

In summary, compared to Baseline Method 1, which executes both the compute-intensive and memory-capacity-bound parts on the CPU, Baseline Method 2 offloads only the compute-intensive part to the GPU. Proposed Method 1 enables both parts to be computed on the GPU through CPU-GPU heterogeneous memory management, and Proposed Method 2 further improves GPU efficiency through memory access management within the GPU memory tiers.

\subsection{Numerical performance}

\begin{figure}[tb]
\begin{center}
\includegraphics[width=\hsize]{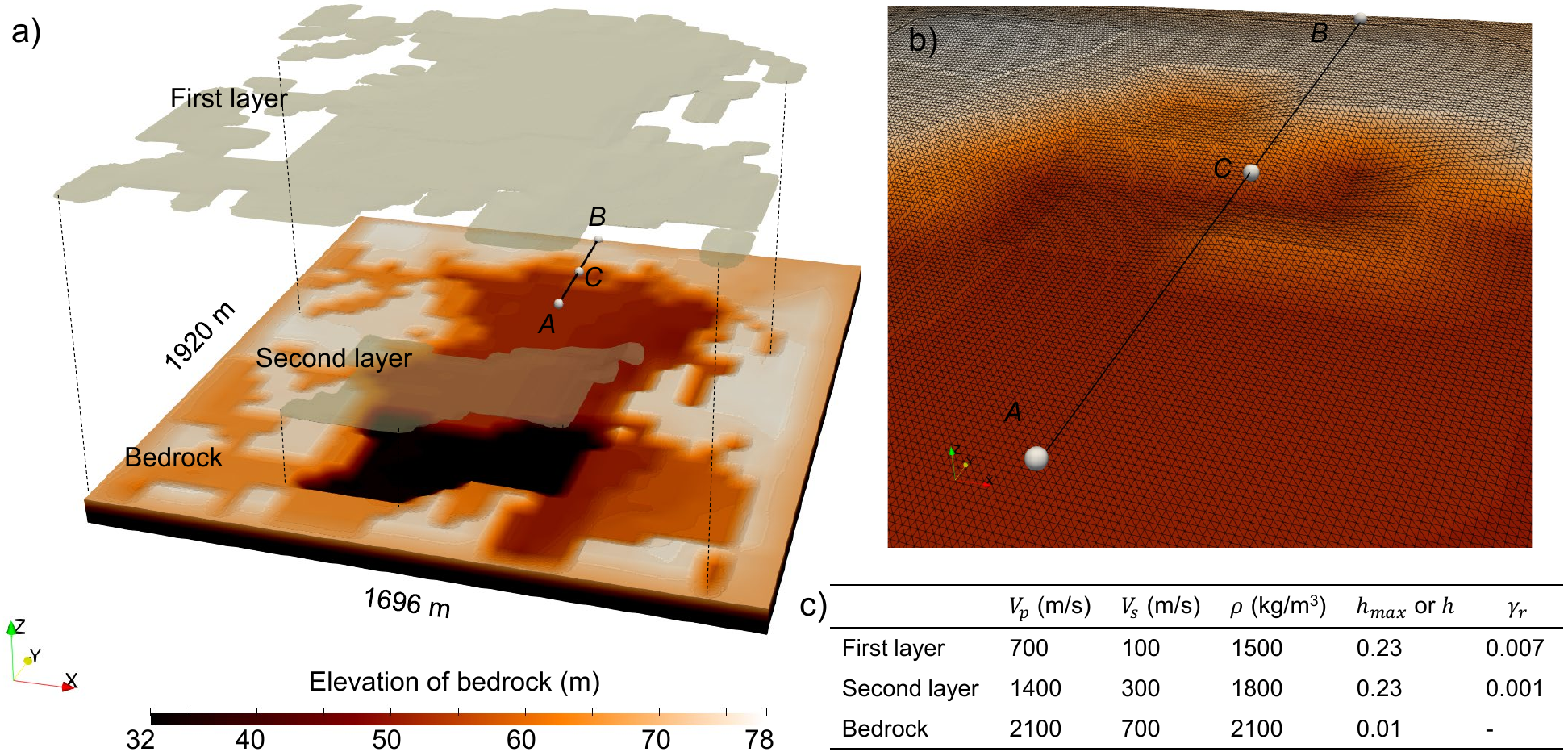}
\end{center}
\caption{{\small (a) 3D ground structure model with line A-B and point C. The $x,y$-coordinates of A, B, and C are (848, 1400), (848, 1900), and (848, 1648) m, respectively. (b) Close-up view of the region around line A-B. (c) Material properties of the soil structure.}}
\label{fig:3Dmodel}
\end{figure}

We utilize a realistic ground model of a site near Tokyo, Japan, which was also employed in \cite{asme2014} (see Fig.~\ref{fig:3Dmodel}). This site is characterized by soft sedimentary layers forming a complex 3D structure, which is known to result in significantly larger ground motions compared to neighboring areas. In other words, it is a site where body waves are converted into surface waves and trapped, leading to pronounced 3D nonlinear ground amplification, making it an ideal model for the performance measurements and application examples presented in this study. In \cite{asme2014}, a comparison was made between observation data from the 2011 Tohoku earthquake -- which caused extensive damage -- and 3D nonlinear analysis results, with the model successfully validated. To analyze the primary components of seismic damage as in \cite{asme2014}, we target a frequency range below 2.5 Hz and generate a finite element model using second-order tetrahedral elements with at least 10 elements per wavelength, following the methodology in \cite{GJI2009}. The model features a minimum element size of approximately 4 m, with 32,502,492 degrees of freedom and 7,781,075 tetrahedral elements. We use a Cartesian coordinate system with the origin at the southwest corner of the model base, where the $x$, $y$, and $z$ axes represent the east-west, north-south, and up-down directions, respectively. The time step increment is set to 0.005 s, consistent with \cite{asme2014}, and the analysis is conducted for 16,000 time steps using the solution methods described in the previous section with a convergence tolerance of a relative error of $10^{-8}$. The input wave for performance measurement is a random wave with frequency components above 2.5 Hz removed, featuring a uniform amplitude distribution ranging from -0.6 to 0.6 for the $x,y$ components and -0.3 to 0.3 for the $z$ component (one of the waveforms used for generating the neural network training dataset in the Application Example).

As an example of a system with high-speed CPU-GPU interconnects, we use the NVIDIA Grace Hopper Superchip (GH200) \cite{GH200} for computational performance measurements. This system is equipped with one 72-core Grace CPU (480 GB memory, 384 GB/s) and one H100 GPU (96 GB memory, 4 TB/s) per node, with the CPU and GPU interconnected via high-speed 900 GB/s NVLink-C2C. Power consumption is measured using the {\tt nvidia-smi -q -d POWER} command to obtain the module power (the entire board including the CPU, CPU memory, and GPU), sampled every 0.5 s and averaged over the total execution time. The power caps per module/GPU are set to the system defaults of 700 W for the GPU and 1,000 W for the module, effectively ensuring no power capping and allowing high simultaneous loads on both the GPU and CPU. The program is implemented using OpenACC for the GPU and OpenMP for the CPU multi-core parts, with 70 cores utilized for OpenMP computation per process. While performance measurements can be implementation-dependent, the implementation in this study is sufficiently optimized to a level comparable to that in \cite{waccpd}. To ensure the baseline methods achieve adequate performance, we employ a $3 \times 3$ Block CRS format to reduce memory access costs. Only the preconditioning part of the solver is computed in single precision, while all other calculations are performed in double precision.

\begin{figure}[tb]
\centering
\includegraphics[width=0.8\hsize]{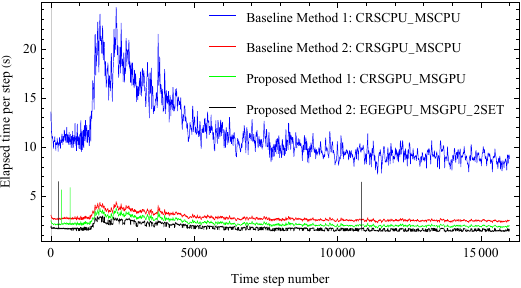}
\caption{{\small Elapsed time per case.}}
\label{fig:elapsedtime}
\end{figure}

Figure~\ref{fig:elapsedtime} shows the elapsed time per case when calculating with each method using one GH200 node. The figure illustrates that near the main motion where the input seismic intensity increases, the convergence deteriorates, leading to an increase in the number of solver iterations and consequently higher execution time per time step, after which convergence improves once the main motion has passed. Table~\ref{tb:performance} presents the total execution time and memory usage required for the entire simulation under these complex and practical problem settings. First, we examine the performance of the conventional baseline methods. As shown in the table, in Baseline Method 1 (CPU-only), the GPU remains idle, resulting in a low average node power of 379 W; however, due to the inability to utilize the high-throughput GPU, the execution time is as long as 182,300 s, and the energy-to-solution is high at 690 MJ. The breakdown of the execution time in Table~\ref{tb:elapsedtime} reveals that the majority of the computational time is consumed by the solver. In Baseline Method 2, where the solver is offloaded to the GPU, the solver time is significantly reduced from 9.4 s to 1.16 s, leading to a reduction in the total execution time by a factor of 4.05. Although the power consumption increases to 635 W due to the use of the high-power GPU, the reduced execution time leads to a decrease in energy consumption to 286 MJ (a factor of 2.41).

\begin{table}[tb]
\centering
\caption{{\small Performance and memory usage of each method. Note that elapsed time and energy usage is shown per case.}}
\small
\label{tb:performance}
\begin{tabular}{cccccc}
\hline
 \multirow{2}{*}{Method}         & Elapsed   & Power & Energy & CPU mem. & GPU mem. \\
                                 & time      & usage & usage  & usage & usage \\
\hline
Baseline Method 1: CRSCPU\_MSCPU   & 182,300 s & 379 W & 690 MJ & 225 GB & -     \\ 
Baseline Method 2: CRSGPU\_MSCPU   &  45,001 s & 635 W & 286 MJ & 131 GB & 66 GB \\ 
Proposed Method 1: CRSGPU\_MSGPU   &  36,074 s & 691 W & 249 MJ & 117 GB & 71 GB \\ 
Prop. Meth. 2: EBEGPU\_MSGPU\_2SET &  14,222 s & 724 W & 103 MJ & 168 GB & 70 GB \\ 
\hline
\end{tabular}
\end{table}

\begin{table}[t]
\centering
\caption{{\small Breakdown of elapsed time of each method. Note that elapsed time is shown per case per time step.}}
\small
\label{tb:elapsedtime}
\begin{tabular}{ccccc}
\hline
\multirow{2}{*}{Method}          & Total  & Solver & CRS  & Multispring total \\
                                 & time   & time   & time & (compute, data transfer) \\
\hline
Baseline Method 1: CRSCPU\_MSCPU & 11.39 s &  9.40 s & 0.92 s & 0.92 s \\ 
Baseline Method 2: CRSGPU\_MSCPU &  2.81 s &  1.16 s & 0.70 s & 0.94 s \\ 
Proposed Method 1: CRSGPU\_MSGPU &  2.25 s &  1.16 s & 0.70 s & 0.38 s (0.33 s, 0.38 s) \\ 
Prop. Meth. 2: EBEGPU\_MSGPU\_2SET& 0.89 s &  0.49 s &      - & 0.39 s (0.34 s, 0.39 s) \\ 
\hline
\end{tabular}
\end{table}

Next, we evaluate the performance of Proposed Method 1, in which the multi-spring calculations are also executed on the GPU. In this method, the entire problem of 7.7 million elements is divided into sub-regions of 0.1 million elements and executed in a pipeline, limiting the increase in GPU memory usage to only 5 GB. Consequently, the entire GPU-utilized portion of the application, including the solver's GPU memory footprint (66 GB), fits within the 96 GB GPU memory limit. The GPU computation and CPU-GPU data transfer for the multi-spring calculation part take 0.33 s and 0.38 s, respectively; as a result, the majority of the computation and transfer can be overlapped, reducing the execution time of the multi-spring part from 0.94 s to 0.38 s compared to the CPU version\footnote{Although the GH200 system supports direct access from within GPU kernels to CPU memory; using this capability increased the multi-spring computation time to 5.9 s due to the longer data-access latency in the NVLink-C2C interconnect, demonstrating the effectiveness of the proposed method.}. This makes the overall application execution time 1.25 times faster than Baseline Method 2 (5.05 times faster compared to Baseline Method 1). As the GPU active time ratio is higher than in Baseline Method 2, the power consumption increases to 691 W, with a total energy consumption of 249 MJ.

Finally, we examine the performance of Proposed Method 2, which features an algorithm designed with a focus on the intra-GPU memory hierarchy. By using the EBE method to compute sparse matrix-vector multiplications on the fly, the need to store massive CRS data is eliminated, reducing GPU memory usage and allowing two problem sets to be solved simultaneously within the 96 GB memory limit. This improves memory access performance and shortens solver time; furthermore, it eliminates the need for CRS update calculations that were previously required every time material properties were updated by the multi-spring model, resulting in an execution time reduction of 1/2.54 compared to Proposed Method 1. Since the power consumption only slightly increases from Proposed Method 1, the energy consumption is reduced to 1/2.42, almost in proportion to the execution time. As a result, we achieve a 12.8-fold speedup (with energy consumption reduced to 1/6.70) compared to the CPU-only Baseline Method 1, and a 3.16-fold speedup (energy reduced to 1/2.78) compared to Baseline Method 2, which only uses the GPU for the solver. Note that if a widely used interconnect such as PCIe Gen 5 x16 (with 1/7 the bandwidth of NVLink-C2C) were employed, the increased data transfer time would outweigh the computational gains of our approach. We can see the usefulness of high-performance hardware combined with suitable algorithm development.

In summary, by increasing the utilization of the high-throughput and power-efficient GPU through heterogeneous memory management, performance is consistently improved across Baseline Method, Proposed Method 1, and Proposed Method 2, in terms of both time-to-solution and energy-to-solution. While these measurements were obtained using a single GH200 module, our approach is expected to scale across large-node systems as the inter-node communication involved is confined to the GPU-based iterative solver, which has already demonstrated high scalability in previous studies.

\section{Application Example}
\label{sct3}

Immediately after a major earthquake, rapid damage assessment is crucial, and there are high expectations for simulations utilizing urban datasets. For instance, \cite{quick} demonstrates a system that estimates input ground motion at the engineering bedrock from observed data, performs 1D nonlinear ground amplification analysis using 1D soil models constructed from urban data at each site, and then calculates the nonlinear time-history dynamic response of individual buildings using the estimated surface motion. However, soil deposits possess three-dimensional structures, which are known to cause local nonlinear ground amplification during large earthquakes -- referred to as 3D dynamic nonlinear effects. To account for these 3D dynamic nonlinear effects, 3D nonlinear analysis using three-dimensional ground structures has proven effective (e.g., \cite{asme2014}), and various attempts have been made to reduce the required computational time. Nevertheless, such large-scale analyses still require significant computational resources and time, as detailed in the previous section, making it impractical to apply them to numerous potential damage sites immediately after an earthquake. Consequently, there is a strong demand for methods that can account for 3D dynamic nonlinear effects while enabling immediate evaluation without intensive computational overhead. In light of this, this section (1) evaluates the 3D dynamic nonlinear effects that can occur during major earthquakes using a realistic 3D ground structure, and (2) demonstrates the utility of the proposed framework by constructing and applying Neural Networks (NNs) that account for these effects using datasets generated from massive ensemble 3D simulations.

Using the ground structure model shown previously (Fig.~\ref{fig:3Dmodel}), we validate the response under nonlinear behavior caused by realistic strong ground motions by inputting the Kobe wave at the bottom of the model and analyzing the ground motions obtained along line A-B and at observation point C. Line A-B corresponds to a location where local ground amplification occurs, and point C is designated as the point on line A-B where the maximum velocity of the $x$-component was recorded during the 3D nonlinear analysis using the Kobe wave. The Kobe wave is based on the ground motion recorded at Nakayamate, Chuo-ku, Kobe, during the 1995 Hyogo-ken Nanbu Earthquake \cite{eqdata}, which caused catastrophic damage, and was prepared using seismic waveforms commonly utilized in Japanese seismic design. Specifically, since this is surface observed data, it was scaled by 1/2 for conversion to an engineering bedrock input wave, and a bandpass filter (0.2-0.5-2.4-2.5 Hz) was applied to focus on the frequency band below 2.5 Hz where damage is concentrated, following \cite{asme2014}. To examine the application performance of NNs for nonlinear problems, we intentionally set a challenging problem where nonlinearity is clearly induced by using somewhat strong, though realistic, ground motion -- achieved by not discounting the ground amplification characteristics during the surface-to-bedrock conversion. The 3D ground amplification analysis in this section employs the same time step, total steps, and convergence criteria as the performance measurements in the previous section. Furthermore, since the evaluation of the waveforms as external forces yielded similar results across the $x$, $y$, and $z$ components, only the $x$-component is presented as a representative case.

\subsection{3D dynamic nonlinear effects}

\begin{figure}[tb]
\begin{center}
\includegraphics[width=0.9\hsize]{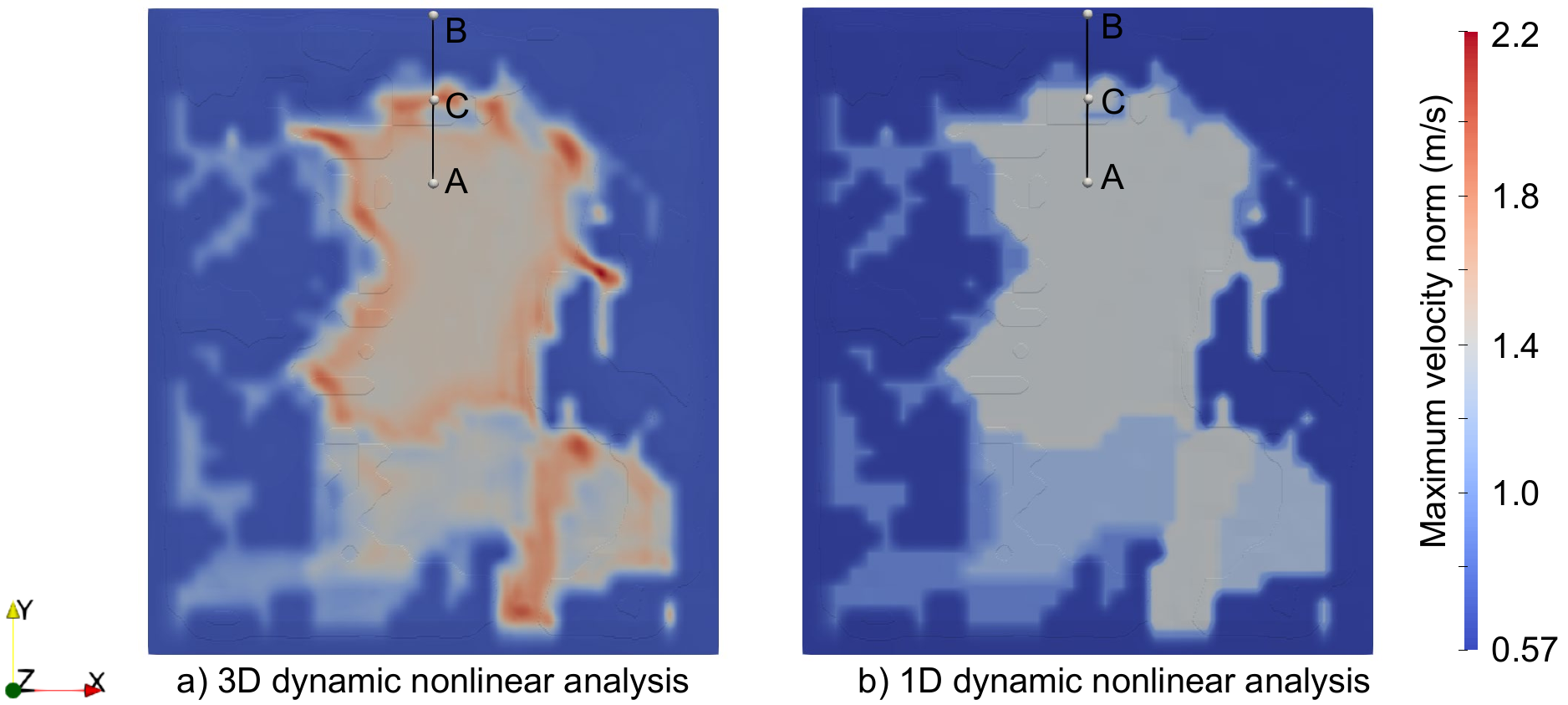}
\end{center}
\caption{{\small Maximum velocity norm distribution at the surface for Kobe wave input
}}
\label{fig:surface_vnorm}
\end{figure}

Figure~\ref{fig:surface_vnorm}(a) shows the maximum velocity norm distribution obtained from the 3D nonlinear dynamic analysis with the Kobe wave input. We can see that local concentrations of ground motion occurs according to the 3D ground structure. Next, Fig.~\ref{fig:surface_vnorm}(b) shows the maximum velocity norm distribution obtained from a 1D nonlinear dynamic analysis with the same Kobe wave input. The 1D nonlinear dynamic analysis, commonly used in ground amplification studies, approximates the soil as a horizontally layered structure, effectively reducing a 3D problem to 1D; while this significantly lowers computational costs, it fails to account for three-dimensional structural complexities. Indeed, comparing Figs.~\ref{fig:surface_vnorm}(a) and (b) reveals significant discrepancies near areas with three-dimensional irregularities.

\begin{figure}[tb]
  \centering
  \begin{subfigure}{0.42\textwidth}
    \centering
    \includegraphics[width=\linewidth]{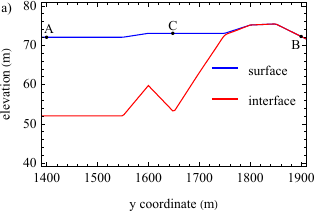}
  \end{subfigure}
  \hspace{0.5 cm}
  \begin{subfigure}{0.42\textwidth}
    \centering
    \includegraphics[width=\linewidth]{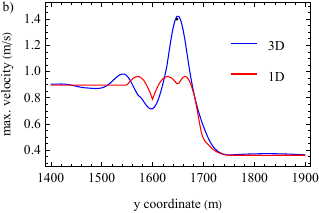}
  \end{subfigure}
  \caption{\small (a) Cross section of ground structure at line A-B. The "surface" indicate the ground surface, while "interface" indicates the interface between the first and bedrock layers. C indicate the obervation point. (b) Maximum velocity response in the $x$-direction along line A-B. The black dot indicates the response estimated by the NNs.}
  \label{fig:crossABVmax}
\end{figure}

\begin{figure}[tb]
  \centering
  \begin{subfigure}{0.43\textwidth}
    \centering
    \includegraphics[width=\linewidth]{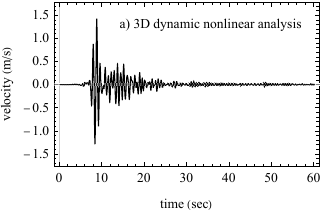}
  \end{subfigure}
  \hspace{0.5 cm}
  \begin{subfigure}{0.43\textwidth}
    \centering
    \includegraphics[width=\linewidth]{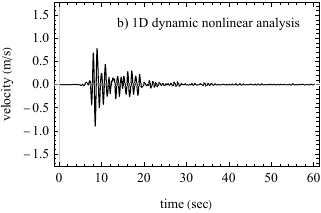}
  \end{subfigure}
  \begin{subfigure}{0.43\textwidth}
    \centering
\includegraphics[width=\linewidth]{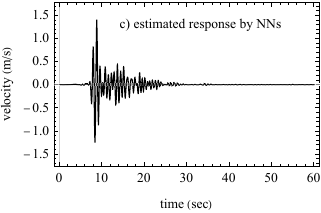}
  \end{subfigure}
  \hspace{0.5 cm}
  \begin{subfigure}{0.43\textwidth}
    \centering
\includegraphics[width=\linewidth]{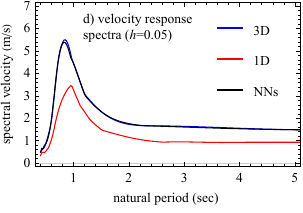}
  \end{subfigure}
\caption{{\small Reponse at point C for Kobe wave. (a) and (b) indicate responses to 3D and 1D dynamic nonlinear analysis, while (c) shows the estimated response by NNs. (d) indicate the velocity response spectra ($h=0.05$) for waves in (a), (b), and (c).}}
  \label{fig:kobewaves}
  \end{figure}

To examine the differences between 3D and 1D analyses in more detail, we check the ground motions along line A-B and at point C shown in Fig.~\ref{fig:3Dmodel}. The ground structure along line A-B possesses an irregularity consisting of the first layer and bedrock with a rising slope, as shown in Fig.~\ref{fig:crossABVmax}(a), which raises concerns about wave concentration (this also includes 3D behavior that cannot be captured by 2D analysis due to non-uniformity in the depth direction). Figure~\ref{fig:crossABVmax}(b) shows the maximum velocity distribution along line A-B for the Kobe wave input. In the 3D analysis, the maximum velocity increases significantly due to local amplification at the slope; however, the 1D analysis fails to capture this, resulting in a significant underestimation. For a more detailed comparison, we examine the waveforms and velocity response spectra at point C. As shown in Figs.~\ref{fig:kobewaves}(a) and (b), the 1D analysis underestimates the waveform amplitude at point C, and its potency as an external force is also markedly underestimated, as evidenced by the velocity response spectrum in Fig.~\ref{fig:kobewaves}(d). These results demonstrate that obtaining time-history waveforms provides more detailed information on external forces useful for damage assessment, and highlights the need for methods capable of accounting for 3D effects where 1D or 2D approximations are inadequate.

\subsection{Effectiveness of NNs considering 3D dynamic nonlinear effects}

We developed NNs capable of estimating time-history waveforms while considering the 3D nonlinear effects of the soil and the characteristics of the input waves, and demonstrate their effectiveness by estimating the waveform at point C. First, 100 random waves were generated with frequency components above 2.5 Hz removed; amplitudes followed a uniform distribution ranging from -0.6 to 0.6 for the $x$, $y$ components and -0.3 to 0.3 for the $z$ component (this setting reflects the fact that vertical components are often smaller than horizontal components in real earthquake records). Using these as input ground motions at the engineering bedrock, we conducted ensemble ground amplification simulations and observed the resulting waveforms at point C. Through this procedure, we obtained 100 sets of time-history data consisting of input random waves ($x, y, z$ components) and the corresponding responses at point C ($x, y, z$ components).

The developed NNs features a symmetric encoder-decoder structure using 1D-CNN and LSTM. In the encoder section ($n_c$ layers), the three input components are temporally compressed while being expanded into a latent dimension $L_{latent}$ to extract local waveform patterns; subsequently, LSTM layers ($n_{LSTM}$ layers) learn temporal features such as nonlinear amplification and delays, after which the decoder section ($n_c$ layers) then predicts the target waveforms from these latent representations. To account for varying physical characteristics among components (e.g., the weaker nonlinearity of the $z$-component), the final layer of the decoder is designed to split the output into three groups for independent convolution. To suppress the excessive influence of peak amplitudes and ensure robust learning of phase and amplitude, we employed MAE loss and the Adam optimizer. Furthermore, using Optuna \cite{optuna}, we optimized the NNs hyperparameters and learning rate $r$ to minimize the validation error. The search space was defined as $n_c \in \{2, 3, 4\}$, $n_{LSTM} \in \{1, 2, 3\}$, convolution kernel size $k \in \{3, 5, 9, 17, 33, 65\}$, $L_{latent} \in \{128, 256, 512, 1024\}$, and $r \in [5 \times 10^{-5}, 5 \times 10^{-4}]$. As a result of the optimization, a model with $n_c=2, n_{LSTM}=2, k=9, L_{latent}=512,$ and $r=1.75 \times 10^{-4}$ (final error: $1.41 \times 10^{-2}$) was selected. The PyTorch-based implementation was run on a single NVIDIA A100 GPU, with the entire training process including the parameter search taking approximately 87 minutes.

Figure~\ref{fig:kobewaves}(c) shows the time-history waveform at point C estimated by the developed NNs for the Kobe wave. It can be seen that the estimation almost perfectly matches the 3D nonlinear analysis results (Fig.~\ref{fig:kobewaves}(a)) and appropriately evaluates the amplification that was missing in the 1D analysis results (Fig.~\ref{fig:kobewaves}(b)). Furthermore, the performance of the waveform as an external force is well-estimated, as shown by the comparison of velocity response spectra (Fig.~\ref{fig:kobewaves}(d)), indicating that the waveforms predicted by the NNs serve as a high-fidelity approximation of the 3D nonlinear analysis results. The difference between the maximum velocity predicted by the NNs at point C and the result from the 3D nonlinear analysis is sufficiently small compared to the order of variation across different locations and input motions, as shown in Fig.~\ref{fig:crossABVmax}(b), confirming the effectiveness of the NN-based evaluation.

\section{Concluding Remarks}
\label{sct4}

In this study, we proposed a novel method based on heterogeneous memory management that enables the execution of massive ensemble simulations for general nonlinear time-history problems with complex constitutive laws by effectively utilizing CPU memory while maximizing GPU computational performance in a CPU-GPU environment. In particular, Proposed Method 2 demonstrated superior performance in terms of both time-to-solution and energy-to-solution, achieving a 12.8-fold speedup (with a 1/6.7 reduction in energy consumption) compared to the CPU-only Baseline Method 1, and a 3.16-fold speedup (with a 1/2.8 reduction in energy consumption) compared to Baseline Method 2, which offloads only the solver to the GPU. In disasters such as earthquakes, highly reliable damage assessments are required immediately after the event; our application example demonstrated that surrogate models constructed using ensemble simulation data enabled by the proposed method can facilitate immediate damage estimation with significantly higher fidelity than conventional approaches. As CPU-GPU interconnect bandwidth and GPU computational speeds continue to increase, the demand for methods like the one proposed here -- which achieve sophisticated simulations by cooperative use of CPU and GPU resources through heterogeneous memory management -- is expected to grow.

\subsection*{Acknowledgments}

The authors thank the Mainline Committee of the Association for the Development of Earthquake Prediction (ADEP) for providing the ground model used in this work. This work was supported by JSPS KAKENHI (25K21686, 23H00213) and supported by JST FOREST Program (JPMJFR215Q).

\end{document}